\newcommand{\PRL}[3]{Phys.\ Rev.\ Lett.\ {\bf #1},\ #2 (#3)}

\newcommand{\NAT}[3]{Nature\ {\bf #1},\ #2 (#3)}
\newcommand{\SC}[3]{Science\ {\bf #1},\ #2 (#3)}

\newcommand{\PRA}[3]{Phys.\ Rev.\ A\ {\bf #1},\ #2 (#3)}

\documentclass[showpacs,preprintnumbers,amsmath,twocolumn,amssymb,pra,superscriptaddress]{revtex4}
\usepackage{latexsym}
 \usepackage{amsmath}
 \usepackage{amsfonts}
 \usepackage{graphicx}
 \usepackage{amssymb}
 \usepackage{dsfont}
 \usepackage{subfigure}

\begin{document}
\title{Grey solitons in doped nonlinear fibers}
\author{Challenger Mishra}
\email{challenger@iiserkol.ac.in}
\affiliation{ Indian Institute of Science Education and
Research Kolkata, Mohanpur 741252, India}

\author{T. N. Dey	}
\email{tarak.dey@iitg.ernet.in}
\affiliation{Department of Physics, Indian Institute of Technology Guwahati, Guwahati 781039, India}

\author{P. K. Panigrahi}
\email{pprasanta@iiserkol.ac.in}
\affiliation{ Indian Institute of Science Education and
Research Kolkata, Mohanpur 741252, India}

\date{\today}
\pacs{42.81.Dp, 42.81.-i,  42.65.-k, 42.65.Hw}

\begin{abstract}
Grey solitons, with complex envelope, are obtained as exact solutions
of the coupled system describing doped optical nonlinear
fibers. The phase degree of freedom plays a crucial role in removing
the restrictive equal frequency conditions on the tanh-sech paired
pulses, identified earlier \cite{Agarwal}. The cross-phase modulation is
found to control the speed of the grey soliton, which has an upper bound. The coupled soliton solution obtained here was found to be stable in a wide parameter range.
\end{abstract}

\maketitle

\section{Introduction}
Self-induced transparency is a dramatic manifestation of nonlinear
effects, leading to coherent localized pulse propagation in an atomic medium \cite{McCallHahn1,EberlyAllen,McCallHahn2,Maimistov}. The controlled, localized
excitation and de-excitation of an ensemble of two level systems is
effectively captured by the hyperbolic secant soliton pulse profile, which propagates as a stable soliton. The Maxwell-Bloch equations
aptly describe the pulse and averaged atomic population dynamics,
which have been extensively investigated for multi-level systems \cite{EberlyAllen,barnard,arecchi,crisp,gsapp,hioe}. Localized and nonlinear cnoidal wave solutions have been
identified, as exact solutions \cite{barnard,arecchi}, which has found experimental
verification \cite{newboldsalamo,shultzsalamo}. 

The subject of pulse propagation in nonlinear
fibers has also been well-studied, since its prediction by Hasegawa
in 1974 \cite{rajupp}. The non-linear Schr\"odinger equation (NLSE), where
the cubic nonlinearity arises due to Kerr effect, describes
the pulse dynamics. This integrable system in one dimension possesses
soliton solutions \cite{zakharovshabat}, which can be dark \cite{burger,deng}, bright
\cite{khaykovich,strecker,khawaja,cornish} or grey \cite{shomroni}, depending on the nature of nonlinearity
and dispersion. Some of these solitons were experimentally observed,
long after their prediction \cite{mollenauer}. Gross-Pitaevskii equation in
one-dimension, which captures the mean-field dynamics of Bose Einstein Condensates (BEC), being identical to NLSE, has also naturally evoked strong interest
in solitons and their dynamics in cigar-shaped BEC \cite{Jackson,pethick}. Dark
solitons in the repulsive regime and bright soliton and soliton trains
in the attractive sector have been experimentally produced \cite{burger,deng,khaykovich,strecker}.
Recently, the grey soliton, having complex profile has been produced
through soliton collision in BEC \cite{shomroni}. Grey solitons have been the subject
of some recent studies \cite{khanatrepp,daspp}. These are analogs of complex envelope
Bloch solitons in condensed matter systems, which can connect differently
ordered domains, without passing through the normal phase. This naturally
makes it energetically favorable, as compared to the N\'eel soliton,
wherein the soliton profile, being of the hyperbolic tangent type,
passes through the normal phase, at the vanishing point of the above
profile. It is interesting to note that the speed of the soliton
can vary from zero to a maximum, depending on the depth of the grey
soliton. Furthermore, the energy shows a maximum value as a function
of momentum. 

The doping of nonlinear fibers with suitable multilevel systems has
opened up the exciting possibility of naturally combining the the atomic system with NLSE, which can possibly lead to effective soliton control
and manipulation. Recently, the localized solitons of a nonlinear
fiber with a three level dopant has been investigated, wherein exact
solitons of $\tanh-\mathrm{sech}$ type have been identified \cite{eberly,rahmaneberly,kozloveberly}.
The constraints of this coupled nonlinear system has led to strong
restrictions on the pulse dynamics. It was found that, the frequencies
have to be perfectly matched for these solutions to exist \cite{Agarwal}. In this paper, we present exact grey soliton profiles in this
system, which removes the above restriction on the pulse
propagation. The complex profile's phase degree of freedom plays a crucial role in relaxing the constraint of frequency
matching. 
\begin{figure}[b]
\includegraphics[width=6.0cm]{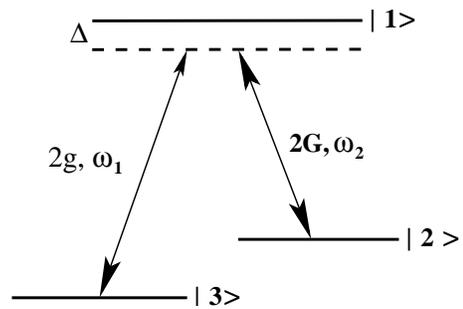}
\caption{\label{Fig1}A three level $\Lambda$ atomic system whose transitions $1\leftrightarrow2$ and $1\leftrightarrow3$ are in near resonance with the two modes of the optical fiber. 
The Rabi frequencies of the two modes are $2G$ and $2g$. $\Delta$ denotes the single photon detuning.}
\end{figure}
\begin{figure}[t]
\includegraphics[width=6.0cm]{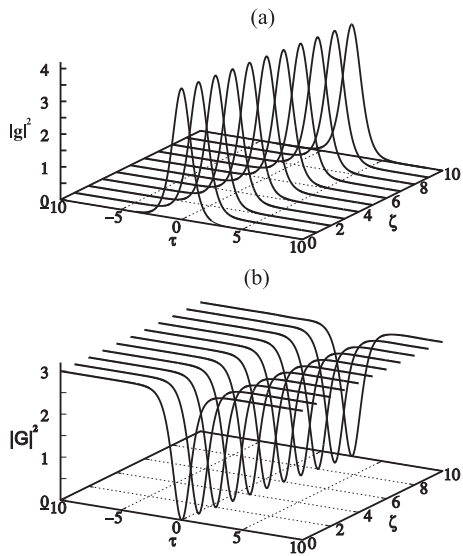}
\caption{\label{Fig2}Stable propagation of (a) bright and (b) dark solitons in a three level medium in
presence of nonresonant nonlinearity and fiber dispersion. This is in the regime of opposite signs of the group velocity dispersion at the two mode
frequencies.}
\end{figure}
\begin{figure}[b]
\includegraphics[width=6.0cm]{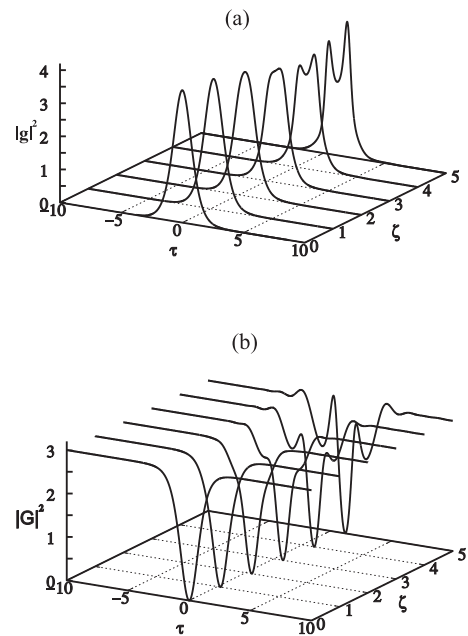}
\caption{\label{Fig3}Growth of instability for (a) bright and (b) dark solitons in a three level medium
in presence of nonresonant nonlinearity and fiber dispersion. This is in the regime of same signs of the group velocity dispersion at the two mode
frequencies.}
\end{figure}

The paper is organized as follows:  In the following, grey solitons, with complex envelope, are obtained as exact solutions
of the coupled nonlinear system describing doped optical nonlinear
fibers. It is explicitly shown how the phase degree of freedom removes the restrictive equal frequency conditions on the $\tanh-\mathrm{sech}$ paired
pulses \cite{Agarwal}. The cross-phase modulation is
found to play a crucial role, in controlling the speed of the grey
soliton.  For the two level system, it is observed that grey soliton is strictly forbidden and only
the SIT soliton can propagate in this dynamical system. Subsequently, we present the numerical techniques and the results of numerical propagation of our solution, which clearly shows stability of the obtained solutions in a wide parameter range. We then conclude with directions for future work.

\section{Coupled grey soliton solutions}

We investigate the possibility of generalization of the solution obtained in \cite{Agarwal} for shape-preserving
pulses in a Kerr nonlinear two-mode
optical fiber doped with 3-level $\Lambda$ atoms as shown in Fig(\ref{Fig1}). For the earlier obtained solutions, the frequency shift was assumed
identical for the two possible fiber modes. The modes of the fiber
are near resonant with the transitions of the atomic system. The two
modes of the optical fiber may be described by the profile,

\begin{equation}
\vec{E}_{i}(z,t)=\vec{A}_{i}e^{-i(\omega_{i}t-k_{i}z)}+c.c.(i=1,2)
\end{equation}

$\vec{A}_{i}$ being the slowly varying envelope, $\omega_{i}$
 and $k_{i}$, the carrier frequency and wave number respectively of the $i^{th}$
mode. The time variation of the population of atomic levels are governed by optical Bloch equations. If $\zeta_{i}(z,t)$
denotes the probability amplitude of the $i^{th}$ atomic level (Fig 1), then the following equations are obeyed within the
rotating wave approximation

\begin{eqnarray}
\dot{\zeta}_{1}&=&-i\Delta\zeta_{1}+iG\zeta_{2}+ig\zeta_{3}\\
\dot{\zeta}_{2}&=&iG^*\zeta_{1}\\
\dot{\zeta}_{3}&=&ig^*\zeta_{1}
\end{eqnarray}

The Rabi frequencies 2g and 2G for the two field modes are related to
the slowly varying amplitudes $\vec{A}_i$, according to the relations

\begin{eqnarray*}\label{Rabi frequency}
2g&=&\frac{2\vec{d}_{13}. \vec{A}_1}{\hbar}\\
2G&=&\frac{2\vec{d}_{12}. \vec{A}_2}{\hbar}
\end{eqnarray*}

where $\vec{d}_{ij}$ is the transition dipole moment matrix element. We take into account the effects of Kerr nonlinearity
(proportional to the square of the electric field) and dispersion,
in order to truly describe the spatio-temporal variation of the light
pulses through the fiber. Using the approximation of slowly varying
envelope one can cast the nonlinear Schr\"odinger equation, in Rabi
frequencies, as
\begin{eqnarray}
\frac{\partial g}{\partial z}&=&-i\beta_{1}\frac{\partial^{2}g}{\partial t^{2}}+i\gamma_{1}(|g|^{2}+2|G|^{2})g+i\eta_{1}\zeta_{3}^{*}\zeta_{1}\\
\frac{\partial G}{\partial z}&=&-i\beta_{2}\frac{\partial^{2}G}{\partial t^{2}}+i\gamma_{2}(|G|^{2}+2|g|^{2})G+i\eta_{2}\zeta_{2}^{*}\zeta_{1}
\end{eqnarray}
where the terms with $\beta_{i}$ are due to group velocity dispersion (GVD), those with $\gamma_{i}$
and $\eta_{i}$ represent the effect of Kerr nonlinearity, and the coupling of the $i^{th}$ mode with the atomic system, respectively. A dark-bright coupled soliton pair was reported in \cite{Agarwal} in this system. There stable propagation of bright and dark solitons was reported, in the regime of opposite signs of the group velocity dispersion at the two mode frequencies. Correspondingly, unstable propagation was reported, when the GVDs were of the same sign. Figs.(\ref{Fig2}) and (\ref{Fig3}) show the solutions obtained by them in stable and unstable regimes of GVD.

\section{Ansatz}

Motivated by the observed grey solitons in BEC \cite{Jackson,khanatrepp},
we assume the following ansatz for the complex envelope solitons
\begin{subequations}
\begin{align}
\label{field_amplitude1}g&=a\cos(\theta)\mathrm{sech}\left[\frac{(t-uz)}{\sigma}\cos(\theta)\right]e^{i(p_{1}z-\Omega_{1}t)}\\
\label{field_amplitude2}G&=b\left[\cos(\theta)\tanh\left[\frac{(t-uz)}{\sigma}\cos(\theta)\right]+i\sin(\theta)\right]e^{i(p_{2}z-\Omega_{2}t)}\\
\label{atom_amplitude1}\zeta_{1}&=\frac{i\alpha\cos(\theta)}{a\sigma}\mathrm{sech}\left[\frac{(t-uz)}{\sigma}\cos(\theta)\right]e^{i(p_{1}z-\Omega_{1}t)}\\
\label{atom_amplitude2}\zeta_{2}&=-\frac{\alpha b\cos(\theta)}{a}\mathrm{sech}\left[\frac{(t-uz)}{\sigma}\cos(\theta)\right]e^{i(p_{1}-p_{2})z-i(\Omega_{1}-\Omega_{2})t}\\
\label{atom_amplitude3}\zeta_{3}&=\alpha\cos(\theta)\tanh\left[\frac{(t-uz)}{\sigma}\cos(\theta)\right]+(1-\alpha)\Gamma
\end{align}
\end{subequations}
\noindent
$u$ gives the envelope velocity in the moving frame. $\sigma$ appearing on the RHS of (\ref{field_amplitude1}) gives the temporal width of the pulse. We have removed the restriction $\Omega_{1}=\Omega_{2}$. The subsequent sections
show that we have been able to derive analogous expressions as in \cite{Agarwal} for the unknown variables introduced in the ansatz.

\section{Consistency conditions}

The ansatz is put into the Bloch and the nonlinear Schr\"odinger
equations (2-6) and the coefficients of $\mathrm{sech}$, $\tanh$, $\mathrm{sech}$$\tanh$,
etc. are collected on both sides to yield the following relations.
The Bloch equations (2-4) give

\begin{subequations}
\begin{align}
\label{consistent equation_a}a^{2}-b^{2}&=\frac{1}{\sigma^{2}}\ ,\\
\label{consistent equation_b}\sin(\theta)&=\sigma(\Omega_{1}-\Omega_{2})\ ,\\ 
\label{consistent equation_c}\alpha&=\frac{a^{2}\sigma^{2}}{a^{2}\sigma^{2}+i(\Delta-\Omega_{1}+b^{2}\sigma\sin(\theta))(\sigma/\Gamma)}\ .
\end{align}
\end{subequations}
The probabilities of occupation of the three levels should add up
to one. Imposing this condition we get,
\begin{equation}
|\Gamma|^{2}=\frac{1-|\alpha|^{2}\cos^{2}(\theta)}{1-|\alpha|^{2}}\ .
\end{equation}
The nonlinear Schr\"odinger equations (5-6) yield
\begin{subequations}
\begin{align}
\label{conp1} p_{1}&=\frac{|\alpha|^{2}\eta_{1}(\Delta-\Omega_{1}+b^{2}\sin(\theta)\sigma)}{a^{4}\sigma^{2}}+2b^{2}\gamma_{1}\nonumber\\
&+\beta_{1}(\Omega_{1}^{2}-\frac{\cos^2(\theta)}{\sigma^{2}})\\
\label{conp2}p_{2}&=\gamma_{2}b^{2}+\beta_{2}\Omega_{2}^{2}\\
\label{conu1}u&=\frac{\eta_{1}|\alpha|^{2}}{a^{2}}+2\beta_{1}\Omega_{1}\\
\label{conu2}u&=\frac{\eta_{2}|\alpha|^{2}}{a^{2}}+2\beta_{2}\Omega_{1}\\
\label{con1}0&=\frac{2\beta_{1}}{\gamma_{1}\sigma^{2}}+(a^{2}-2b^{2})\\
\label{con2}0&=\frac{2\beta_{2}}{\gamma_{2}\sigma^{2}}+(2a^{2}-b^{2})
\end{align}
\end{subequations}
Equation (\ref{conu2}) can be alternately cast as,
\begin{equation}
\label{group velocity}
u=2\beta_{2}\Omega_{2}+(b^{2}-2a^{2})\gamma_{2}\sigma\sin(\theta)+\eta_{2}\frac{|\alpha|^{2}}{a^{2}}
\end{equation}
For stable propagation of pulses $\beta_{1}\ne\beta_{2}$ \cite{Agarwal}. The following constraint can be obtained from equations (\ref{conu1}) and (\ref{conu2})
\begin{equation}\label{constraint1}
\frac{|\alpha|^{2}}{a^{2}}(\eta_{1}-\eta_{2})+2\Omega_{1}(\beta_{1}-\beta_{2})=0\ {\textrm {or}}\ \Omega_{1}=\frac{\frac{|\alpha|^{2}}{a^{2}}(\eta_{2}-\eta_{1})}{2(\beta_{1}-\beta_{2})}\\
\end{equation}
Equations (\ref{consistent equation_a}), (\ref{con1}) and (\ref{con2}) lead to (\ref{constraint2}), which is an important result that
needs to be satisfied by the nonlinearity and dispersion in the medium
at the two mode frequencies:

\begin{equation}\label{constraint2}
3+2\frac{\beta_{1}}{\gamma_{1}}+2\frac{\beta_{2}}{\gamma_{2}}=0
\end{equation}

Another important point to note is that, for $\theta=0$, in order to have a solution, $\Omega_1= \Omega_2$ is required. This is because of the consistency relation  (\ref{consistent equation_b}) viz., $\sin \theta= \sigma\big(\Omega_1 - \Omega_2\big)$. If $\Omega_1 \ne \Omega_2$, then $\sigma=0$ and the arguments of all the sech and tanh functions in the ansatz (\ref{field_amplitude1}-\ref{atom_amplitude3}) blow up to give constant solutions. Therefore $\Omega_1 = \Omega_2$ is required when $\theta=0$. Thus in \cite{Agarwal}, while looking for pure dark soliton solution for G, the condition $\Omega_1 = \Omega_2$ was automatically enforced. In comparison, in our solution, we fix $\Omega_2$ to any value and find the rest of the parameters in terms of $\theta$ ($\theta\ne0$, $\Omega_1 \ne \Omega_2$). Therefore our ansatz (\ref{field_amplitude1}-\ref{atom_amplitude3}) represents valid solutions for all $\theta$, given the consistency conditions are satisfied, showing that grey solitons of varying depths can be solutions of the system. 

\begin{figure}[t]
\noindent \centering{}
\includegraphics[scale=0.45]{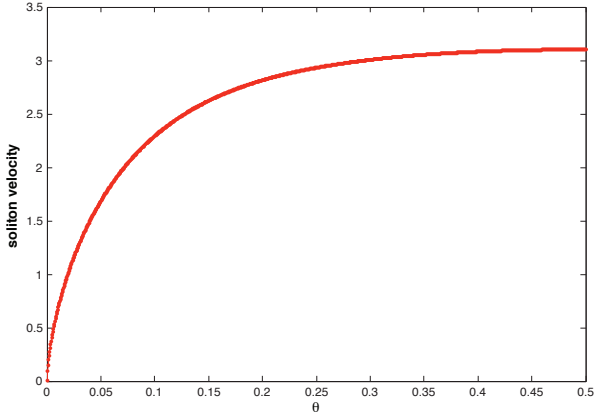}
\caption{\label{fig4}Grey solitons are usually velocity restricted. Often the relationship is as simple as $u\propto\sin\theta$. In our case it is given by (\ref{group velocity}). The parameters used are  $\beta_1=1$,
$\beta_2=-2.5$, $\gamma_1=1$, $\gamma_2=1$, $\eta_1 = 1$, 
$\eta_2 = 1.2$ and $\Delta=1$.}\end{figure}

The velocity of the solitons (u) has a nontrivial dependence on $\theta$. This physically means that the velocity of the solitons is dictated by the depth of the grey soliton in the system. Fig (\ref{fig4}) shows that the solitons are velocity restricted with maximum value attained at $\theta=\pi/2$.

\section{Stability of the coupled grey and bright soliton pair}

\subsection{Linear stability analysis}

Linear stability analysis is done by perturbing around the known solutions. Perturbing around 5 complex waveforms results in 10 coupled differential equations in 10 variables with complicated potentials (composed from the solutions themselves), which are difficult to handle either analytically or numerically. Hence, we have carried out numerical evolution of the pulse for the purpose of stability check. Here, we state the equations of motion of the perturbations, as derived from the equations of motion for $g$, $G$  and $\zeta_i$ (2-6) by substituting $g\rightarrow \big(\tilde{g}+\delta \tilde{g}\big)e^{i(p_{1}z-\Omega_{1}t)}$, $G\rightarrow \big(\tilde{G}+\delta \tilde{G}\big)e^{i(p_{2}z-\Omega_{2}t)}$, $\zeta_1 \rightarrow \big(\tilde{\zeta_1}+\delta\tilde{\zeta_1}\big)e^{i(p_{1}z-\Omega_{1}t)}$, $\zeta_2 \rightarrow \big(\tilde{\zeta_2}+\delta\tilde{\zeta_2}\big)e^{i(p_{1}-p_{2})z}e^{-i(\Omega_{1}-\Omega_{2})t}$ and $\zeta_3 \rightarrow \big(\tilde{\zeta_3}+\delta\tilde{\zeta_3}\big)$. The quantities $\tilde{g}$, $\tilde{G}$ and $\tilde{\zeta_i}$ are given by the RHS of equations (\ref{field_amplitude1}-\ref{atom_amplitude3}) respectively, without the exponential terms. The perturbations satisfy,

\begin{widetext}
$\left[\begin{array}{ccccc}
\Gamma_{11} & \Gamma_{12} & \eta_{1}\tilde{\zeta_3}^{*} & 0 & \eta_{1}\tilde{\zeta_1}\mathrm{conj}\\
\Gamma_{21} & \Gamma_{22} & -\eta_{2}\tilde{\zeta_2}^{*} & -\eta_{2}\tilde{\zeta_1}\mathrm{conj} & 0\\
-i\tilde{\zeta_3} & -i\tilde{\zeta_2} & \Gamma_{33} & -i\tilde{G} & -i\tilde{g}\\
0 & -i\tilde{\zeta_1}\mathrm{conj} & -i\tilde{G}^{*} & \Gamma_{44} & 0\\
-i\tilde{\zeta_1}\mathrm{conj} & 0 & -i\tilde{g}^{*} & 0 & \partial_{t}\end{array}\right]\left[\begin{array}{c}
\delta \tilde{g}\\
\delta \tilde{G}\\
\delta\tilde{\zeta_1}\\
\delta\tilde{\zeta_2}\\
\delta\tilde{\zeta_3}\end{array}\right]=0$
\end{widetext}

where the operators $\Gamma_{ij}$ stand for:

\begin{subequations}
\begin{align}
\Gamma_{11}&=-i\big(\partial_{z}+ip_{1}\big)+\beta_{1}\big(\partial_{t}^{2}-2i\Omega_{1}\partial_{t}-\Omega_{1}^{2}\big)\nonumber\\
&-2\gamma_{1}\big(|g|^{2}+|G|^{2}\big)-\gamma_{1}g^{2}\ \mathrm{conj}\\
\Gamma_{22}&=-i\big(\partial_{z}+ip_{2}\big)+\beta_{2}\big(\partial_{t}^{2}-2i\Omega_{2}\partial_{t}-\Omega_{2}^{2}\big)\nonumber\\
&-2\gamma_{2}\big(|g|^{2}+|G|^{2}\big)-\gamma_{1}G^{2}\ \mathrm{conj}\\
\Gamma_{33}&=\partial_{t}-i\Omega+i\Delta\\
\Gamma_{44}&=\partial_{t}-i\big(\Omega_{1}-\Omega_{2}\big)\\
\Gamma_{12}&=-2\gamma_{1}gG^{*}-2\gamma_{1}gG\ \mathrm{conj}\\
\Gamma_{21}&=-2\gamma_{2}g^{*}G-2\gamma_{2}gG\ \mathrm{conj}
\end{align}
\end{subequations}

Here, $\mathrm{conj}$ is the function that takes the complex conjugate of the term it acts on eg., $\mathrm{conj}(\delta \tilde{g})=\delta \tilde{g}^*$. The above matrix equation is thus a condensed form of an actual set of 10 coupled partial differential equations in 10 variables.

\begin{figure}[t]
\includegraphics[width=8.0cm]{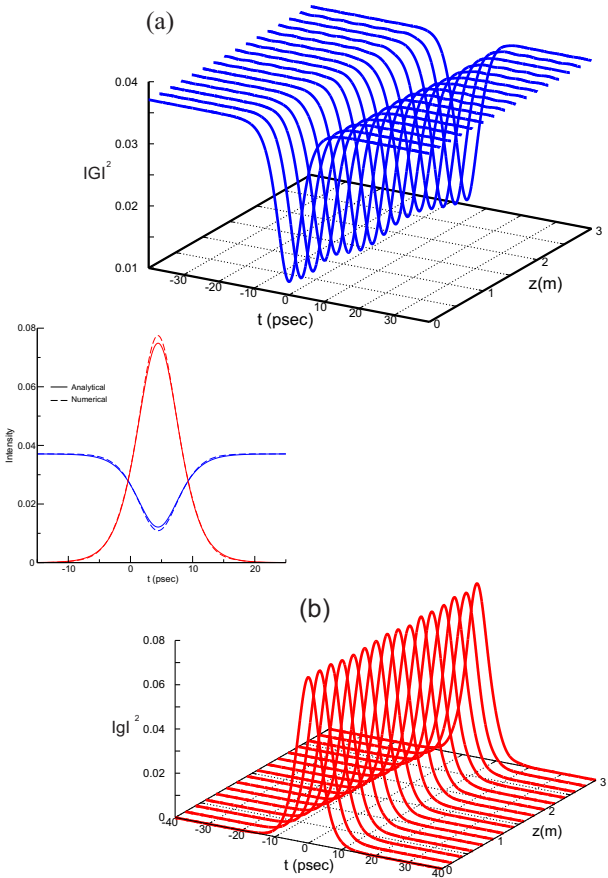}
\caption{\label{fig5}Spatio-temporal evolution of the grey-bright soliton pair in medium  plotted against  the spatial length of the medium at different propagation times. 
The other parameters used are A=0.33361, B=0.19261, $p_1=1.0039$, $p_2=0.037098$, $\Omega_1=0.15584$, $\Omega_2=0$, $\theta=0.60908$, $\beta_1=-0.25$, $\beta_2=-1.25$, $\gamma_1=1$, $\gamma_2=1$, $\Delta=1$.  $u = 1.4805$,  $\sigma= 3.6712$, $\Gamma=1.0338$, $\eta_1=1$, and $\eta_2=1.2$.}
\end{figure}

\begin{figure}[t]
\includegraphics[width=7.0cm]{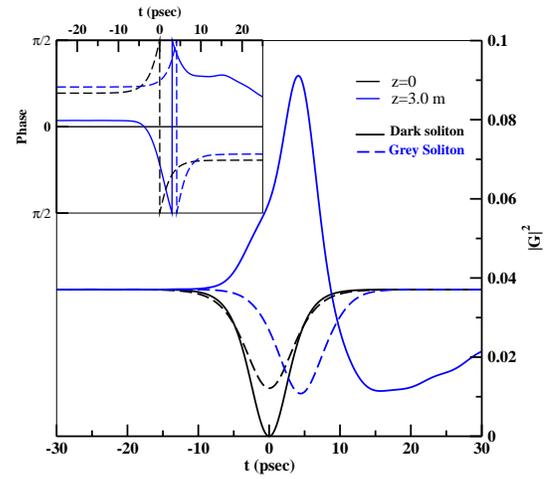}
\caption{\label{fig6}Amplitude and phase of the dark and grey solitons are plotted as a function of time at the exit of the medium. All other parametrs are as in Fig. \ref{fig5}.}
\end{figure}
\begin{figure}[b]
\includegraphics[width=6.0cm]{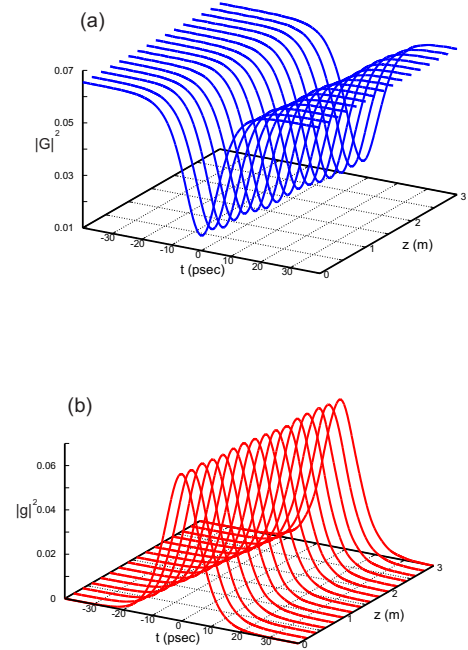}
\caption{\label{fig7}Spatio-temporal evolution of the grey-bright solitons using
parameters such that the group velocity dispersion have the opposite sign. 
We used A=0.29551, B=0.25592,$p_1=0.77581$, $p_2=0.065494$,$\Omega_1=0.072505$, $\Omega_2=0$, $\theta=0.51291$, $\beta_1=1.0$,
$\beta_2=-2.5$ ,$\gamma_1=1$, $\gamma_2=1$, $\Delta=1$.  $u = 2.6827$,  $\sigma= 6.768$, $\Gamma=1.0337$, $\eta_1=1$, and $\eta_2=1.2$.}
\end{figure}

\subsection{Numerical stability analysis}
In this section, we investigate the propagation dynamics of the grey and bright solitons by integrating the full set of coupled nonlinear Sch\"odinger and Bloch equations.  We use Runge-Kutta and split step operator methods to investigate the spatio-temporal evolution of the optical pulses in fiber doped with three level $\Lambda$ system. The initial shapes of the input pulse and the initial atomic populations  can be obtained in the limit of $z=0$ and $t=0$ from the analytical expressions (\ref{field_amplitude1}-\ref{field_amplitude2})  and (\ref{atom_amplitude1}-\ref{atom_amplitude3}) respectively. 
We use  grey and bright solitons with width $3.6712$ psec in the present investigation. Parameters used in the numerical simulation obey self consistent relations (\ref{constraint1}) and (\ref{constraint2}).  We first explore the stable propagation of the grey and bright solitons in presence of  GVD with same character. Figure (\ref{fig5}) shows that the amplitudes of the output pulses are retaining their initial shape. Both solitons propagate through the medium with same group velocity.  A small oscillation though has been noticed at both the ends of the grey soliton due to the modulation instability.  Our analytic and numerical solution of grey-bright soliton pairs are presented in the inset of the figures (\ref{fig5}) at the propagation length $z=3$m. The numerical results obtained from the Runge-Kutta and split step operator methods match well with those obtained analytically. Next we study how the blackness parameter  $\theta$ leads to stable propagation of soliton through the medium even though GVD has same sign. 
A comparative study of the dark and grey solitons through the medium is shown in the figure \ref{fig6}. The dark soliton can be obtained with $\theta =0$ whereas $0<\theta<\pi/2$ represents the grey soliton. For dark soliton, the intensity of the dip always vanishes but the same for the grey soliton always remains. The variation of phase with time with propagation length for both dark and grey solitons is seen to be antisymmetric in nature. Interestingly, dark soliton shows a sudden change in phase across the line center while grey soliton displays no change in shape of phase across the line center during the length of propagation. In the course of propagation, the antisymmetric phase has a temporal shift due to the presence of self and cross phase modulations of the medium in both the cases. The dark soliton develops instability, both in intensity as well as in the phase while grey soliton propagates through the medium without loss of generality upto sufficient length of propagation. Hence the instability due to the same GVD sign can be compensated with the proper choice of the blackness parameter $\theta$.

Figures (\ref{fig7}a)-(\ref{fig7}b) show the results for the spatio-temporal evolution of grey and bright solitons in presence of the opposite group velocity dispersion.  Both the grey and bright solitons maintain their initial shape.

\section*{Conclusions}
In conclusion, grey solitons, analog of Bloch solitons in condensed
matter systems, have been identified as exact solutions of the coupled
nonlinear Schr\"odinger and Bloch equations. The complex nature of
the profile played a crucial role in removing the restrictive frequency
matching condition of $\tanh-\mathrm{sech}$ paired solutions \cite{Agarwal}. We numerically evolved our solution and found it to be stable in both regimes of group velocity dispersion.  We note that, for the case of coupling with a two-level system ($g\rightarrow0$, $\zeta_3\rightarrow0$) instead of the $\Lambda$ system considered here, the pure $\tanh$p	 and grey solitonic profiles are prohibited for G, while a pure sech profile is allowed. The fact that our solutions are stable for some parameters in both regimes of group velocity dispersion signs, is very interesting and needs further investigation.

\paragraph*{Acknowledgement:} PKP would like to thank Prof. G. S. Agarwal for discussions and many useful inputs.

\end{document}